# Distributed Adaptive Formation Control for Multi-UAV to Enable Connectivity

Rahim Rahmani*, Ramin Firouzi, Theo Kanter
Department of Computer and Systems Sciences
Stockholm University
Nod Building, SE-164 55 Kista, Sweden

**Abstract**

There is increasing demand for control of multi-robot and as well distributing large amounts of content to cluster of Unmanned Aerial Vehicles (UAV) on operation. In recent years several large-scale accidents have happened. To facilitate rescue operations and gather information, technology that can access and map inaccessible areas is needed. This paper presents a disruptive approach to address the issues with communication, data collection and data sharing for UAV units in inaccessible or dead zones and We demonstrated feasibility of the approach and evaluate its advantages over the Ad Hoc architecture involving autonomous gateways.

## 1. Introduction

In the smart urban environments and mobile sensing computing application usually require an adaptive control of community sensing, ubiquitous connectivity, open data distributed processing and decision making are creating new smart community connection to reduce traffic congestion, to provide public safety and make local government more efficient more and more urban collection areas are beginning to harness the power of wearable sensors, mobile devices with build-in sensing abilities and other connected devices and engage citizens equipped with smart devices. An issue strogly related to design, development and control of such system needs a distributed control of cooperation multi autonomous vehicles which has been extensively studied in the last. The problem of design development and control of multi agent systems consisting of multiple autonomous robots or Unmanned Aerial Vehicles (UAV) has been proposed in order to meet the requirement of complex missions [1]. We believe that assigning multiple UAV to perform tasks for autonomous cooperative decision-making in Distributed task systems are critical in order to realize distributed adaptive control for multi-UAV as autonomous cooperative enabled decision-making system.

In order to meet the demanding challenges facing autonomous multiple UAV networks in real application in this paper we discuss fundamental connections between distributed control, planning, perception and decision making for multiple UAV networks and propose 3 the following key technical challenges, Control & cooperative for Distributed Adaptive system, Communication and Cooperative localization. The concept shows autonomy in UAV management in distributed multiple UAV systems. As shown in Figure 1 in response to real-time monitoring condition multi-UAV can be dispatched for some typical monitoring tasks over different disaster evacuation. The rest of the paper is structured as follows. Section 2 defines the background and related works of Distributed Adaptive Formation Control for multi UAV and Section 3 describes the conceptual model of the framework; Section 4 presents the performance evaluation and the conclusions are provided in section 5.

## 2. Background and Related Works

In this section we briefly discuss the background and our motivation behind this work and related works. As an important of many applications of urban sensing and environmental monitoring uses sensor to collect data. However, the share of data a fundamental need to accomplish multiple tasks across space and time that are beyond the capabilities of a single autonomous platform. The growing of demand of distributed control of Multi-UAV has stimulated a broad interest in distributed adaptive data processing strategies that support for control & cooperative of Multi-UAV as well as UAV autonomy. To support efficient autonomous multiple UAV networks in a scenario described in Figure 1 in real application two main challenges must be addressed.

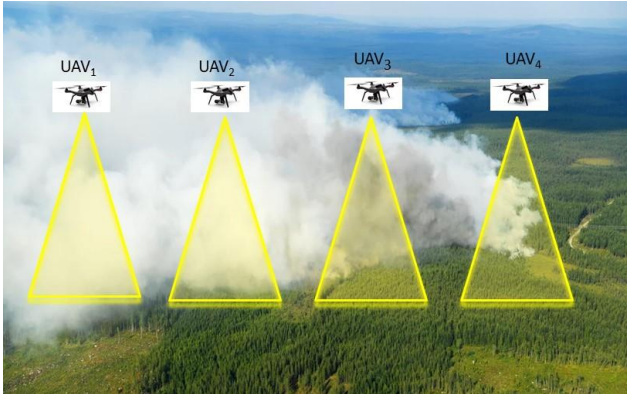

Figure 1. Multi-UAV view of the mission scenario

## 2.1 Adoption of Distributed Multiple UAV

These limitations mandate of centralized cooperative strategies to operate as an alternative approach intelligently moving control to the "things" to better adapt to the link status for intra / inter multi-UAV data dissemination. So the question becomes how to extend adoption of distributed multiple UAV for the capability of each UAV to be able to avoid unpredicted the obstacles in the airspace and at the same time guarantee a minimum separation distance to the other UAV in cooperation mission. In the [2], [3] had proposed motion planning and control and adaptive formation control of Multi-UAV by orthogonal transformation addressed in [4]. In addition, some online and adaptive control and cooperative with centralized and decentralized strategies haven been proposed in [5]-[8]. However, none of these works have proposed a feasible scheme for adoption of the link status for intra / inter multi-UAV data dissemination.

## 2.2 Control & cooperative for Distributed Adaptive system

One of main property of an autonomous operation framework is to address cooperative mission planning, coordinated motion control and collision avoidance. The research on autonomous operation multi robot have been addressed in [6], [8]. However, none of these research work addressed the issue of how control and cooperative of multi-UAVs mission safety which is in the context of our proposed approach. To increase the in-filed time of a multi-UAV as monitoring grid that collect heterogenous data, we had to design an adaptive optimization method to cooperatively monitoring area such as cooperative path following of multi-UAV.

We will discuss the specific design of architectural design to address above challenges in the following sections

## 3. System Architecture Design

### 3.1 Distributed control of multiple autonomous UAV

Existing approaches which in principle allow us to move control to end-points such as edge computing and mobile ad-hoc networking rely on various routing and addressing principles. In contrast, we require a comprehensive end-to-end reachability mechanism across autonomous [16] Multi-UAV. In fact consensus can be considered a special case of formation control. Consensus, a fairly basic problem in distributed multi-UAV coordination. Rules-based framework is one of milestone of distributed coordination. Rules of engagement are one of the important aspects of IoT that allows collected data to provide more meaning and intelligence to the raw data. There have been many rule-based techniques available employing rule-engine for consensus- based solutions. The autonomous UAV will enable creation and maintenance of rules of engagement of the UAVs. Consensus based rules will make communication [20] and distributed task assignment protocols to be fault tolerant or resilient. Processes using the rules will be able to communicate with one another, and agree on a single consensus value. Distributed control in end-nodes must be able to negotiate distributed coordination problems, it is, sometimes, assumed that some global information is available to each individual UAV. This assumption disobeys the virtue of distributed multi-UAV coordination. As an alternative, distributed control methodologies have been proposed in which some unknown global information can be estimated locally. Existing approaches such as the edge computing lack a single concrete architecture supporting these requirements. Multi-UAV nodes can make decisions based on context at two levels in response to a) the unavailability of communications paths and b) application level events and context changes [18]. Finally, we require interworking between the multi-UAV, which in essence, is a peer-to-peer IoT infrastructure [19], [21] and a cloud-based IoT platform hosted in a data-center. In support of delegation from a cloud-based IoT platform we designed and implemented an interworking function which is able to interpret scripted instructions from a cloud platform and return the results after delegating it to the swarm of IoT nodes. In this section we explain architectural design of the system in details. First we describe the system model as shown in figure 2. Second, we present the autonomic of the UAV and its relation to multi-UAV infrastructure.

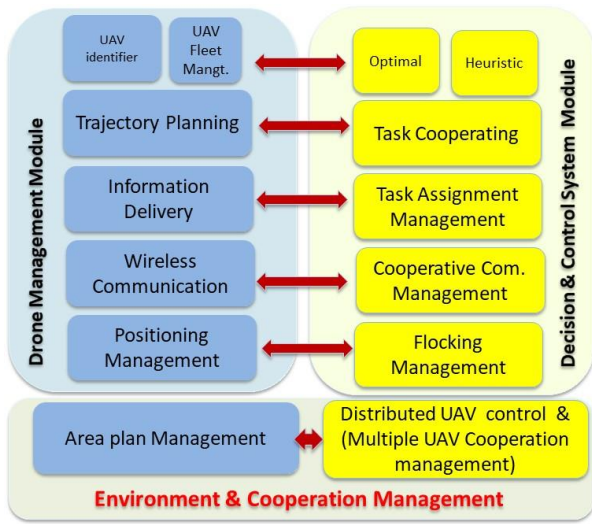

Figure 2. System Model Architecture

3.2 The system Model

The components of the UAV control system architecture illustrated in figure 2 including UAV management module, Environment & cooperation management module and decision and control module. The functionality of Drone management module is to detect population and other UVA thus to form the distribution of information. Environment & cooperation management module responsibility is to manage area plan and multiple UAV cooperation as a cluster [17]. The functionality of decision and control module is an online adaptive learning for an adaptively communication between UAVs. The workflow of autonomic UAV and online adaptive learning shown in figure 3. In the system framework the integration of these control modules and movement-planning can be summarized:

- Positioning and flock management algorithms to produce a set of spatial proximity, paths and profile that account for UAV dynamics and spatial coordination constraints and inter-UAV safety requirements.

- Trajectory planning and Task cooperating algorithms that enables the UAVs to follow the paths while adjusting the speed of the UAVs to ensure coordination.

- Cooperative communication algorithm that enable UAVs to exchange data and forwarding UAVs sensor data to base station during flight. In the use-case which shown in figure 1 it is important to have an accurate and up-to-date overview of the situation. Some of areas are more interest while others are minor interest. Each observation area has certain quality parameters assigned (e.g., spatial and temporal data). During mission execution the overview area is assigned through the Task assignment management to the UAVs identifier and UAVs fleet and incrementally refined and update as the mission advances. Interesting objects with in the observation areas are highlighted. The UAV can adapt the observation areas according to the current situation.

3.3 Autonomic UAV

Autonomic UAV is designed to be use as part of massive adoption of distributed UAV. There will be some form of autonomous device coordination and a single UAV arbiter of roles and permission such a solution grants greater power to the owners of UAV to define how UAV interact via rules of engagement. Rules of engagement are one of the important aspects of distributed adaptive control for Multi-UAV that allows collected data to provide more meaning and intelligence to the raw data. Online learning is implemented in the approach to periodically and adaptively balance rate and loss rate. The workflow of autonomic UAV and online adaptive learning shown in figure 3. Given the UAVs scenario definition the three main steps performed by the system for generating an overview monitoring and scanning are 1) planning the mission 2) executing the mission and 3) analysing the data.

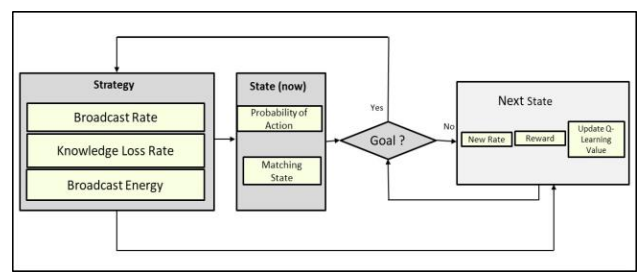

Figure 3. Workflow of Adaptive Formation Control

The UAVs scenario definition serves as the input for the planner. UAV together with sensors and its online learning (Q-learning) capabilities the first step is to compute the positions by learning the environment and evaluating an action value function which gives the expected reward of taking an action in a given state the distributed learning agent is able to make a decision automatically. The next step is to compute routes for the UAVs so that each time step the agent is given a reward than new values are calculated for each combination of a state from the set of states and action from the set action while minimizing the energy consumption of each UAV and distributing the workload equally. The core of the algorithm is a simple value iteration update make possible for UAVs which fly individually or in formation sensing environment and coordination of UAVs, UAVs flying in formation and the impact of communication. During flight the UAVs collect data at the planned observation area and the run-time of online learning process is illustrated in state transition diagram in figure 3. The process state includes the broadcast rate and residual energy. To meet Multi-UAV mission accuracy requirements as confident event logs have particularities that are important in model discovery. The impact of multiple tasks poses a difficulty for discovery algorithms.

## 4. Performance Evaluation

In this section we evaluate the performance of the proposed system architecture through experiments & scenarios. Using a ex-post evaluation we employed the system architecture in several experiments to determine if the system lives up to our defined requirements and functional.

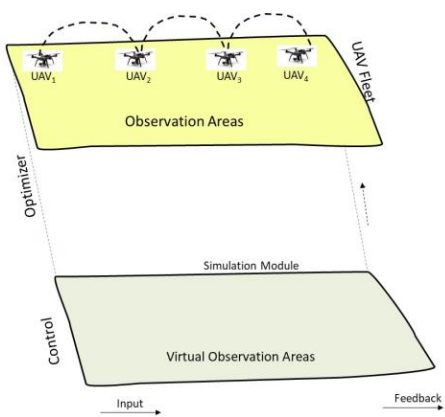

Figure 4. Overview of the UAV control system in action when several UAVs are coupled together to move in separate areas.

### 4.1 Multi-UAV Environment and System Architecture

The environment management and control is illustrated as shown in figure 4. In this experiment include two scenarios. One scenario in which we updated in order to ensure that the simulation module can reinstate the UAVs communication if the UAVs within the observation area would for some reason lose connection and the UAVs control system will receive notification as inputs for adjusting observation area for the UAVs. The UAV fleet [9] first scanning and monitoring around the observation areas.

The objective of this scenario was to experiment the system architecture ability to reestablish connections with devices that are temporarily disconnected due to some interference such as obstacles which is related to requirement such as if the system architecture should consider obstacles such as mountains or might disrupt the link chain. The data in table 1 was collected by measuring how long it took for the control system to reestablish connection between all UAVs after a disruption. For each distance the simulation was performed five times and the data is an average of all the data collected from the simulations.

Table 1: Mean time for UAVs to re-establish connection after disconnection

| Distance | Elapsed time to reestablished connection | |
|---|---|---|
| | Device 1 and Device 2 | Device 2 and Device 3 |
| 4m | 5,44% | 12,620s |
| 8m | 7,7452s | 10,7166s |
| 12m | 6,017s | 10,181s |

In second scenario when a UAV would fly to close to another UAV, which would result in the tracks overlapping and redundant data collection, the control system should be able to sense this and instruct the UAV to move out of the other UAVs track. This scenario was created to test the adaptability of the control system in such an event. This scenario also tests the accuracy of the distance estimation which is specified as a requirement (accuracy of at least 80% on distance estimation), where accuracy is how accurately a device is able to estimate the distance of another UAV by using Received Signal Strength Indicator (RSSI) [10] values. The accuracy is measured by how often the RSSI values stay within an RSSI interval. This scenario has been logged with measurements of RSSI values for every four meters from 4 meters up to 16 meters. The goal of this scenario has been to measure how well the UAVs can stay within parameters specifying the device to stay at the respective distances.

Table 2: Statistics value of second scenario to estimate the distance by using RSSI

| Meters | RSSI Interval | Device 1 | | | Device 2 | | |
|---|---|---|---|---|---|---|---|
| | | Successes | Failures | Accuracy | Successes | Failures | Accuracy |
| 4m | -2 <-> -8 | 0 | 513 | 0,00% | 258 | 257 | 50,10% |
| 4m | -3 <-> -7 | 0 | 513 | 0,00% | 216 | 328 | 39,37% |
| 4m | -2 <-> -6 | 0 | 513 | 0,00% | 196 | 348 | 36,03% |
| 8m | -10 <-> -15 | 260 | 258 | 50,19% | 139 | 355 | 28,14% |
| 8m | -11 <-> -14 | 166 | 352 | 32,05% | 63 | 431 | 12,75% |
| 8m | -12 <-> -13 | 68 | 450 | 13,13% | 18 | 476 | 3,64% |
| 12m | -14 <-> -20 | 309 | 198 | 60,95% | 385 | 124 | 75,64% |
| 12m | -15 <-> -19 | 242 | 265 | 47,73% | 232 | 277 | 45,58% |
| 12m | -16 <-> -18 | 138 | 369 | 27,22% | 75 | 434 | 14,73% |

The control system was not able to perform this scenario well, which shown in figure 5 as we can see the accuracy is at interval -14 to -20 RSSI for both UAV 1 (60,95%) and UAV 2 (75,64%), though the requirement was 80%. Neither does the system architecture fulfill requirement because the RSSI measurement was very volatile. Bluetooth RSSI is susceptible to noise and as shown in table 2 the measured RSSI was not able to stay within specified ranges very well or at all. This makes it very hard for the UAV to sense each other accurately within the observation areas. This issue has been identified as a hardware issue since the system architecture itself would have performed well if the measurements collected from the bluetooth dongle had been more stable and resistant to noise. This become apparent when looking at all the table 2 and figure 5 detailing RSSI data where two UAVs measured differently to each other. However, the system architecture does fulfill communication and connection requirements where a UAV can send messages to its neighbours if they are too close or too far away.

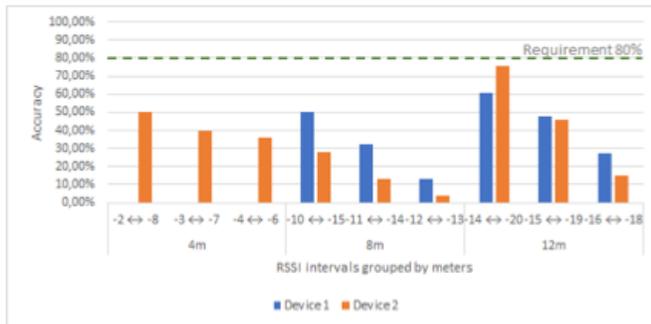

Figure 5. Accuracy for UAV1 and UAV2 by RSSI intervals of distance

## 4.2 Experiment of Scanning & Monitoring Observation Area via UAVs Localization by BreezySlam Algorithm

The definition of requirements has in this experiment has been based on document studies about previous research on the Simultaneous Localization and Mapping (SLAM) algorithm [11] and assumptions made in regards to the characteristics of the system architecture. The process began at the beginning of the experiment and continued during so the requirements has in an iterative way been changed or added. This was made because new requirements had to be added by the control system when problems during the development were encountered. The requirements are mainly functional in nature and touches upon essential functionality and possibilities of the simulation tool, the UAVs and laser scanner and the BreezySlam algorithm [12]. The simulation tools were the programs in where an UAV equipped with a laser scanner and its behaviour was simulated to produce data that is processed by the SLAM-algorithm [13], [14] and [15].

In this experiment include two scenarios. One scenario in which an environment builds by several obstacles where an UAV equipped with a laser maps the observation area as it moves through it with a velocity of 0.2 m/s as shown in figure 6 (a). The path was closed, which means the UAV after going a full lap ends up where it started. The UAV turns with an angle of 90°, since the laser scanner always must face the motion direction.

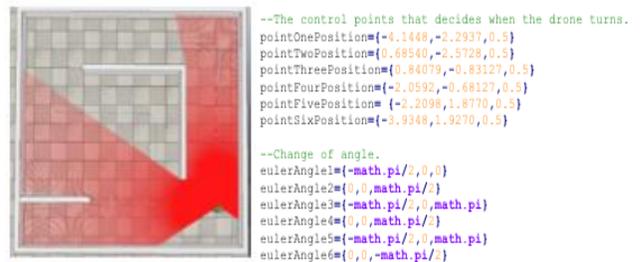

Figure 6. a) Observation area path b) The cod controlling the UAV turning 90º.

To made the UAV turn when it should, in the curves, the control positions for the change of the angle and the angles itself shown in cod figure below. The parameters that controlled the behaviour of the UAV and the laser scanner during the simulation of scenario 1shown in table 3.

Table 3 Simulation Parameter for Scenario 1

| Parameter name | Value |
|---|---|
| Simulation time | 00:1:34:95 (dt=50.0ms) |
| Length of the laser scanner | 5 meters |
| Scope of the Laser scanner | 240º |
| Scanning Frequency | 10 Hz |
| Size of scan | 684 |
| Max velocity of drone | 0.2 m/s |
| Max acceleration of drone | 0.1 m/s |

Through observation of the imaging of scenario 1 shown in figure 7 and the mapping the of simulated environment shown in figure 8, can several differences be seen. Some extra obstacles have been added, like the one in the lower left corner, and some were missing, like the one at the top. Furthermore, the path of the UAV that the algorithm has calculated and imaged as most probable, see the black curvy line in figure 7 compared to the arrow in the same figure, was also erroneous.

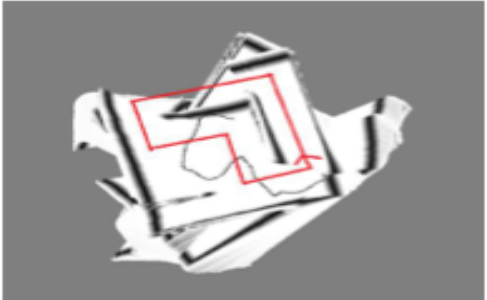

Figure 7. The Simulation result of scenario 1 with estimation path generated by BreezySlam

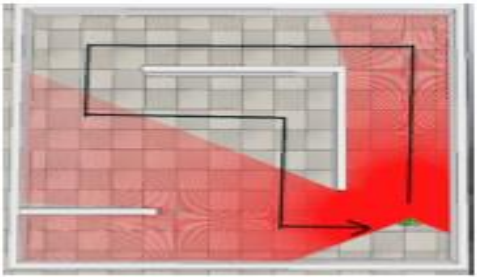

Figure 8. The path in Scenario for th eUAV

Second scenario contains more obstacles and corridors compared to the first ones and the UAV was standing still as shown in figure 9. The purpose of this scenario was compare the result of the mapping with a drone that moves and maps at the same time, and to test the artefact's ability deal with the mapping of more objects. Since the UAV was standing still, no additional code for motion planning has been written.

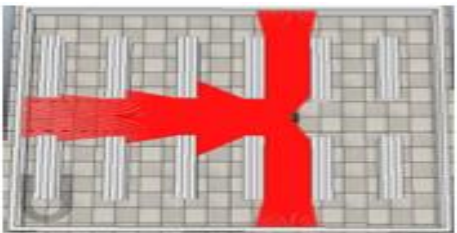

Figure 9. The path in Scenario 2 for the UAV

The parameters that controlled the behaviour of the UAV and the laser scanner during the simulation of scenario 2 shown in table 4. Through observation and comparison of figure 10 (a) and figure 10 (b) the conclusion could be drawn that the algorithm in an almost correct way have imaged that which the laser scanner has mapped during the simulation of scenario 2. The objects that are missing are the ones that the laser scanner senses at 0° and 240° respectively.

Table 4. Simulation parameter for scenario 2

| Parameter name | Value |
| --- | --- |
| Simulation time | 00:0:05:00 (dt=50.0ms) |
| Length of the laser scanner | 5 meters |
| Scope of the Laser scanner | 240° |
| Scanning Frequency | 10 Hz |
| Size of scan | 684 |
| Max velocity of drone | 0.0 m/s |
| Max acceleration of drone | 0.0 m/s |

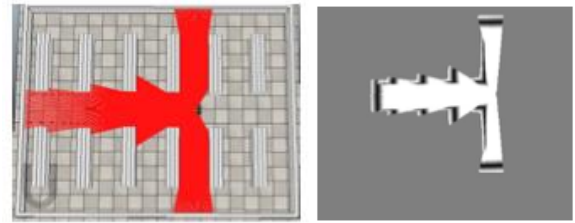

Figure 10. a) The simulated Environment that was mapped in scenario 2.   b) The result of the imaging of scenario 2

## 5. Conclusions

The true promise of comprising multiple UAVs for the application in disaster management can only benefit smart community sensing when a middleware solution for adaptive distributed data processing and dissemination for multiple UAVs in disaster area in order to control & cooperative of Multi-UAV as well as UAV autonomy. The work presented an autonomy concept in UAV management in distributed multiple UAVs systems in response to real-time monitoring condition and multi-UAVs can be dispatched for some typical monitoring tasks over different disaster evacuation. Our approach extends an enable connectivity for multiple UAVs and mapping of surrounding observation areas with SLAM-algorithm. Thereby shows distributed control in end-nodes must be able to negotiate distributed coordination problems, it is assumed that some global information is available to each individual UAV. This assumption disobeys the virtue of distributed multi-UAV coordination. Experiment results shows distributed control methodologies on unknown global information can be estimated locally.  Further results show Multi-UAV nodes can.

   make decisions based on context in response to the unavailability of communications paths and application level events and context changes.

The work presented in this chapter can be extended further. The proposed adaptive & learning approach can be implemented and its feasibility can be investigated. Currently, we are working towards employing deep learning on drones to learn from the scanning and monitoring observation areas. It's feasibility to improve Smart scanning is one of the future works we are focusing currently, specially in disaster management and monitoring plant-growth at the edge to improve observation and management of disaster areas. In the future, interoperability and dynamic behaviour between UAVs controllers, and edge and cloud controller interoperability should be examined. Intelligence and security are the two outstanding issues in the current and future IoT. Therefore, security in IoT needs to be explored from the distributed multiple UAVs intelligence perspective.